\documentclass[10pt]{article}
\usepackage{latexsym}
\usepackage{amssymb}
\usepackage{amsfonts}
\usepackage{amsmath}
\usepackage{theorem}

\renewcommand{\o}{_{\rm o}}

\def\be{\begin{equation}}
\def\ee{\end{equation}}
\def\bc{\begin{center}}
\def\ec{\end{center}}

\begin{document}

\title{Spin glass polynomial identities from entropic constraints}
\author{Peter Sollich\footnote{King's College London, Department of Mathematics, Strand, London WC2R 2LS, U.K., peter.sollich@kcl.ac.uk}, Adriano Barra\footnote{Dipartimento di Fisica, Sapienza Universita' di Roma, P.le Aldo Moro 5, 00185, Roma, Italy,  adriano.barra@roma1.infn.it}}
\maketitle

\begin{abstract}
\noindent The core idea of stochastic stability is that thermodynamic observables must be robust under small (random) perturbations of the quenched Gibbs measure. Combining this idea with the cavity field technique, which aims to  measure the free energy increment under addition of a spin to the system, we sketch how to write a stochastic stability approach to diluted mean field spin glasses which explicitly gives overlap constraints as the outcome. We then show that, under minimal mathematical assumptions and for gauge invariant systems (namely those with even Ising interactions), it is possible to ``reverse'' the idea of stochastic stability and use it to derive a broad class of constraints on the unperturbed quenched Gibbs measure. This paper extends a previous study where we showed how to derive (linear) polynomial identities from the ``energy'' contribution to the free energy, while here we focus on the consequences of ``entropic'' constraints. Interestingly, in diluted spin glasses, the entropic approach generates more identities than those found by the energy route or other techniques. The two sets of identities become identical on a fully connected topology, where they reduce to the ones derived by Aizenman and Contucci.
\end{abstract}

\section{Introduction}

Polynomial identities have a long history in spin glass theory, from the early development by Ghirlanda and Guerra (GG) \cite{gg} and by Aizenman and Contucci (AC) \cite{ac,guerra2} at the end of the 1990s. The link with the peculiar organization of states (in the low temperature phase) discovered by Parisi \cite{MPV} was guessed immediately; however, it is only in the past few years -- and only for fully connected mean field systems, namely the Sherrington-Kirkpatrick model (SK) \cite{sk} -- that Panchenko has been able to show the deep connection between polynomial identities and ultrametricity \cite{panchenko1,panchenko2,panchenko3}.

Following the seminal approaches, the former based on checking the stability of states by adding all possible $p$-spin terms \cite{gardner,derrida,grem} and then sending their strength to zero, the latter using a property of robustness of the quenched Gibbs measure with respect to small stochastic perturbation \cite{ac,pierluz,contucci}, identities for the SK model have by now been obtained with a number of different techniques, e.g.\ via smooth cavity field expansion \cite{barra1}, linear response stability \cite{claudio1}, random overlap structures \cite{peter} or even as Noether invariants \cite{BG1}.
In the diluted counterpart of the SK model, which is the Viana-Bray model (VB) \cite{vb}, a similar research effort has produced classes of identities that naturally generalize the ones obtained earlier by AC (see for instance \cite{barra4,franz1}) and GG (see for instance \cite{franz2,t1}).

It has since been possible to show to validity of these polynomial identities even in short-range, finite-dimensional models \cite{boviernuovo,claudio2,t1,T}, and novel techniques to obtain other identities, with the aim of finding a set of constraints on the overlap probability distribution strong enough to enforce the replica symmetry breaking scheme, are still of great interest, especially  beyond the SK framework (see e.g.\ \cite{claudio3,parisitala}).

In this paper, to complement the analysis begun in \cite{peter} (where we showed how to obtain polynomial identities in spin glasses by considering the energy contribution of random overlap structures developed in \cite{ass}), we show how to derive AC-like polynomial constraints for the Viana-Bray model even from the entropic contribution. Namely, we add a (random) perturbation term in the Boltzmann factor -- close in spirit to the cavity field approach -- and then show that, in the thermodynamic limit, this is irrelevant on average as it coincides with a negligible shift in the connectivity. However, the introduction of this ``innocent'' perturbation, within the standard stochastic stability framework, enables us to derive linear combinations of the constraints on the perturbed Boltzmann measure. The latter converges to the unperturbed measure and returns the identities (all together, combined into an infinite series) as a consequence.

To obtain the constraints as separate identities, we go further and ``reverse'' the idea of stochastic stability. We introduce the random perturbation only as an overlap generator via derivatives; once we then get the desired polynomials, we evaluate all their averages within the original unperturbed quenched Gibbs measure. Remarkably this procedure, subject to minimal mathematical assumptions, produces separately all the AC-like identities (which reduce to the standard AC constraints in the SK model limit of high connectivity), generalizing all known results.  Our arguments do not amount to a rigorous proof, but we hope that they may serve as inspiration for future work in this direction.

\section{Model, notations, cavity perspective and preliminaries}

In this section we provide a streamlined summary of previous results to make the paper self-contained. In particular, after introducing the model (and the associated standard statistical mechanics definitions), we explain in two further subsections the cavity field and stochastic stability perspectives as applied to diluted spin glasses, with the aim of showing the deep link between these two approaches. In the last subsection we discuss a decomposition of the free energy that highlights the synergy among the cavity field and stochastic stability points of view, and provides a suitable starting point for our investigation of polynomial overlap constraints.

\subsection{The diluted spin glass}

To introduce the model (originally studied by Viana and Bray in \cite{vb}), let us consider $N$ Ising spins $\sigma_i\in\{+1,-1\}$, with $i$ running from 1 to $N$; $\sigma=(\sigma_1,\ldots,\sigma_N)$ will denote the complete spin configuration.
Let $P_\zeta$ be a Poisson random variable of mean $\zeta$,
and let the $\{J_\nu\}$  be independent and identically distributed copies of a random coupling strength variable
$J$ with symmetric distribution. For the sake of simplicity, and
without undue loss of generality \cite{gt2}, we will assume $J=\pm 1$.
The coupling strengths $J_\nu$ will determine binary interactions between spins at sites $\{i_\nu\},\{j_\nu\}$; the latter are
independent identically distributed random variables, with uniform
distribution over $1,\ldots,N$.
If there is no external field,
the Hamiltonian of the Viana-Bray (VB) model for dilute mean
field spin glass is then
\begin{equation}
\label{ham}
H_N(\sigma, \alpha; \mathcal{J})=
-\sum_{\nu=1}^{P_{\alpha N}} J_\nu \sigma_{i_\nu}\sigma_{j_\nu}\ ,\
\alpha\in\mathbb{R}_+\ .
\end{equation}
The non-negative parameter $\alpha$ is
called {\sl degree of connectivity}: if the sites $i$ are regarded as
vertices of a graph, and the pairs $(i_\nu,j_\nu)$ define the edges of
this graph, then each vertex is the endpoint of on average $2\alpha$
edges as explained below.

The Hamiltonian (\ref{ham}) as written has the advantage that it
is the sum of (a random number of) i.i.d.\ terms. To see the
connection to the original VB-Hamiltonian, note that the
Poisson-distributed total number of bonds obeys $P_{\alpha
N}=\alpha N + O(\sqrt{N})$ for large $N$. As there are $N^2$
ordered spin pairs $(i,j)$, each gets a bond with probability
$\sim \alpha/N$ for large $N$. The probabilities of getting two,
three (and so on) bonds scale as $1/N^2,1/N^3,\ldots$ so can be
neglected. The probability of having a bond between any unordered
pair of spins is twice as large, i.e.\ $2\alpha/N$. For large $N$
each site therefore has on average $2\alpha$ bonds connecting to
it, and more precisely this number of bonds to each site has a
Poisson distribution with mean $2\alpha$.
The self-loops that we have allowed just add $\sigma$-independent constant to the Hamiltonian so are irrelevant.

We will denote by $\mathbb{E}$ the expectation
with respect to all the (quenched) variables,
i.e.\ all the random variables
except the spins, collectively denoted by $\mathcal{J}$. The Gibbs measure $\omega$ is defined by
$$
\omega(\varphi)=\frac{1}{Z_N(\alpha,\beta)}\sum_{\sigma}
\varphi(\sigma) e^{-\beta H_N(\sigma, \alpha; \mathcal{J})}
$$
for any observable $\varphi:\{-1,+1\}^{N}\to \mathbb{R}$,
where $Z_N(\alpha,\beta)= \sum_{\sigma}\exp(-\beta H_{N}(\sigma,\alpha;\mathcal{J}))$ is the partition function for a finite number of spins $N$.
When dealing with more than one configuration, the product
Gibbs measure will be denoted by $\Omega$, and spin
configurations taken from each space in such a product are called
``replicas''. We use the symbol $\langle . \rangle$ to mean $\langle . \rangle = \mathbb{E}\Omega(.)$.

We will often omit the dependence on
the size of the system $N$ of various quantities, a convention already deployed above. In general, we will allow slight
abuses of notation to lighten the expressions as long as there
is no risk of confusion. The pressure $P_N(\alpha,\beta)$ and the free
energy density $f_{N}(\alpha,\beta)$ for given system size $N$ are defined by
$$
P_N(\alpha,\beta)= -\beta f_{N}(\alpha,\beta)=\frac1N\mathbb{E}\ln Z_{N}(\alpha,\beta),
$$
and we assume that the limit $\lim_{N \to \infty}P_N(\alpha,\beta)=P(\alpha,\beta)$ exists.

The entire physical behavior of the model is encoded by
the distribution of the (even) multi-overlaps $q_{1\ldots 2n}$, which are functions of
several configurations $\sigma^{(1)},\sigma^{(2)},\ldots,\sigma^{(2n)}$ defined by
\begin{equation*}
q_{1\ldots 2n}=\frac1N \sum_{i=1}^N
\sigma_i^{(1)}\cdots\sigma_i^{(2n)}\ .
\end{equation*}
By studying the behavior of these order parameters it is possible to obtain a phase diagram for diluted spin glasses in the $(\alpha,\beta)$ plane which consists of an ergodic phase (where all overlaps vanish in the thermodynamic limit of large $N$) and a spin glass phase (where the overlaps are positive), separated by a second order critical line given by
\be
2\alpha \tanh(\beta)=1.
\ee

\subsection{The cavity perspective}

Following the idea at the heart of the cavity approach (namely, measuring
the effect on the free energy of the addition of one spin to the system;
see \cite{ass2,peter} for a summary), we write, in distribution,
\begin{equation}\label{step}
H_{N+1}(\sigma, \sigma_{N+1}, \alpha; \mathcal{J}) =
-\sum_{\nu=1}^{P_{\alpha \frac{N^{2}}{N+1}}}
J_{\nu}\sigma_{i_{\nu}}\sigma_{j_{\nu}}
-\sum_{\nu=1}^{P_{\alpha \frac{2N}{N+1}}}
J'_{\nu}\sigma_{i'_{\nu}}\sigma_{N+1}
-\sum_{\nu=1}^{P_{\frac{\alpha}{N+1}}}
J''_{\nu}\sigma_{N+1}^2 \ ,
\end{equation}
where $\sigma_{N+1}$ is the added spin. The $\{J'_{\nu},J''_\nu\}$ are independent copies of
$J$, and $\{i_{\nu}\}$, $\{j_{\nu}\}$, $\{i'_{\nu}\}$ are independent
random variables all uniformly distributed over $\{1,\ldots,
N\}$. The last term in (\ref{step}) does not contribute when $N$ is
large, and at any rate is a constant which cancels from the Boltzmann measure.

Note that we can equivalently write the above decomposition as
\begin{equation}\label{acca1}
H_{N+1}(\sigma, \sigma_{N+1}, \alpha; \mathcal{J}) = H_{N}(\sigma,\alpha^{\prime};\mathcal{J})
+h_{N+1}\sigma_{N+1}
\end{equation}
where
$$\alpha^{\prime}=\alpha\frac{N}{N+1}\ ,\
h_{N+1}=-\sum_{\nu=1}^{P_{2
\alpha^{\prime}}}J'_{\nu}\sigma_{i'_{\nu}}\ .
$$
Exploiting the additivity property of Poisson variables, we can also decompose the Hamiltonian for an $N$-spin system so that it shares the first term with $H_{N+1}$:
\begin{equation}\label{acca2}
H_{N}(\sigma,\alpha;\mathcal{J})=H_{N}(\sigma,\alpha^{\prime}; \mathcal{J})
+H_{N}(\sigma,\alpha^{\prime}/N; \mathcal{\hat{J}})\ ,
\end{equation}
where the two Hamiltonians on the right hand side have
independent quenched random variables $\mathcal{J}$ and $\mathcal{\hat{J}}$.
Hence, if we call
$$
H_{N}(\sigma;\alpha^{\prime}/N; \mathcal{\hat{J}})
=\hat{H}_N(\sigma,\alpha^{\prime};\mathcal{\hat{J}})=
-\sum_{\nu=1}^{P_{\alpha^{\prime}}}\hat{J}_{\nu}
\sigma_{\hat{i}_{\nu}}\sigma_{\hat{j}_{\nu}}\ ,
$$
then
$$
\mathbb{E}\ln\frac{Z_{N+1}(\alpha,\beta)}{Z_{N}(\alpha,\beta)}
=\mathbb{E}\ln\frac{\sum_{\sigma,\sigma_{N+1}}
\xi_{\sigma}\exp(-\beta h_{N+1}\sigma_{N+1})}{\sum_{\sigma}
\xi_{\sigma}\exp(-\beta \hat{H}(\sigma,\alpha^{\prime};\mathcal{\hat{J}}))}\ ,
$$
with
$$
\xi_{\sigma}=\exp(-\beta H_N(\sigma,\alpha^{\prime};\mathcal{J}))\ .
$$
As elegantly explained in \cite{ass2}, and discussed in detail in \cite{peter}, this equation expresses
the incremental contribution to the free energy in terms of the
mean free energy of a spin added to a reservoir
whose internal state is described by $(\sigma, \xi_{\sigma})$,
corrected by an inverse-fugacity term $\hat{H}$,
which encodes a connectivity shift.
The former
may be thought of as the {\em cavity} into which the
$(N + 1)$ particle is added:
for $N \gg 1$, the value of the added
spin, $\sigma_{N+1}$, does not significantly affect the
field that would act for the next increment in $N$.
Hence, for the next addition
of a particle we may continue to regard the state of the
reservoir as given by just the
configuration $\sigma$. However, the weight of the
configuration (which is still to be normalized
to yield the probability of the configuration) changes according to
$$
\xi_{\sigma}\to\xi_{\sigma}e^{-\beta h_{N+1}\sigma_{N+1}}\ .
$$
This transformation is called {\sl cavity technique}.

\subsection{The link to stochastic stability}

The
addition of a new spin can, because of the randomness of the couplings,
effectively be regarded as an external random field that vanishes in the thermodynamic limit. This is essentially the perspective of the stochastic stability approach \cite{pierluz,contucci}.

By  an interpolation method \cite{barra1} the ($N+1$)-th spin can be
added to the $N$-spin system smoothly via an appropriately defined cavity
function $\Psi(\alpha,\beta,t)$, $t \in [0,1]$, which reads
\begin{equation}\label{psi2}
\Psi(\alpha,\beta,t) =\mathbb{E}\ln\omega(e^{\beta \sum_{\nu=1}^{P_{2\alpha t}}
J'_{\nu}\sigma_{i'_{\nu}}}).
\end{equation}
Due to the gauge symmetry of the VB model, namely the symmetry $\sigma_{i_{\nu}}\to \sigma_{i_{\nu}}\sigma_{N+1}$ (whose action leaves the VB Hamiltonian invariant), the above cavity function turns out to
contain an effective two body interaction, as in the original Hamiltonian,
and the sum over $\sigma_{N+1}=\pm 1$ in the partition function gives a trivial factor two because $\sigma_{N+1}$ plays the role of a
hidden variable; this factor two yields, once the logarithm is
taken, just the high temperature entropy.

Inspired by the cavity perspective, taking $\varphi$ as a generic function of the spin configuration, we can define
a generalized Boltzmann measure (denoted by the subscript $\langle
. \rangle_t$) as
\begin{equation*}
\omega_t(\varphi)=\frac{\omega(\varphi(\sigma) e^{\beta\sum_{\nu=1}^{P_{2\alpha
t}}J'_{\nu}\sigma_{i'_{\nu}}})}{\omega(e^{\beta\sum_{\nu=1}^{P_{2\alpha
t}}J'_{\nu}\sigma_{i'_{\nu}}})}. \end{equation*}
Note that in the $t =0$ case we always recover the
unperturbed Boltzmann measure of an $N$-spin system and in the
$t =1$ case we recover the unperturbed Boltzmann measure of
an $N+1$-spin system, with a small shift in the connectivity that
becomes negligible in the thermodynamic limit.

Let us now  briefly
describe the stochastic stability properties for
averaged overlap correlation functions (OCFs); these will become useful shortly.
We split OCFs into two categories: filled OCFs, showing {\itshape
robustness} with respect to the stochastic perturbation, and
fillable OCFs, showing {\itshape saturability} with
respect to the same perturbation.
\begin{itemize}
\item Filled OCFs are monomials in overlaps among $s$ replicas
such that each replica appears an even number of times.
Examples are $q_{12}^2$, $q_{1234}^2$ or $q_{12}q_{23}q_{13}$.

\item
Fillable OCFs are overlap monomials among $s$ replicas
which become filled when multiplied by a single overlap among exactly
those replicas appearing only an odd number of times.
Examples are  $q_{12}$, $q_{1234}$
or $q_{12}q_{13}$.
\end{itemize}
It should be pointed out that all monomial OCFs are either filled or fillable, because one can always find a multioverlap to fill any (monomial) OCF that is not filled. This contrasts with the case of the SK model, where only overlaps among two replicas can be used to fill an OCF \cite{barra1}.
The division into filled and fillable OCFs is made because of differences in how their averages react to the perturbing field induced by the cavity function \cite{barra1,barra4,peter}. In the thermodynamic limit, the averages of the filled OCFs become independent of $t$, i.e.
$$
\lim_{N \to \infty}\partial_t \langle \mbox{filled\ OCF}\rangle_t = 0. $$
We refer to this property as {\itshape robustness}.

On the other hand, using the gauge symmetry, one has in the thermodynamic limit that the averages of fillable OCFs become filled at $t=1$, namely \cite{barra1,barra4,peter}
$$
\lim_{N \rightarrow \infty} \langle \mbox{fillable\ OCF}
\rangle_{t=1} = \lim_{N \rightarrow \infty} \langle \mbox{filled\ OCF} \rangle_{t=1}= \lim_{N \rightarrow \infty} \langle \mbox{filled\ OCF} \rangle.$$
We refer to this last property as {\itshape saturability}. Note that
we have dropped the subscript $t$ in the last equality because of the
robustness of filled OCFs. Examples of saturability are $\langle q_{12} \rangle_{t=1}
 = \langle q_{12}^2 \rangle$, $\langle q_{1234}\rangle_{t=1} = \langle q_{1234}^2 \rangle$ and $\langle q_{12}q_{13} \rangle_{t=1}  = \langle q_{12}q_{13}q_{23} \rangle$, with the limit $N\to\infty$ always understood.

We only sketch the proof of the above propositions and refer the reader to
\cite{barra1,barra2,barra4,peter} for a detailed discussion
and proofs.
Let us show how the fillable OCFs turn out
to become filled OCFs in the $N\rightarrow \infty$ limit at $t=1$.
The
stability of the filled OCFs will then be a straightforward
consequence of their gauge invariance, which is heavily used in the proof.
Consider the simplest case of
a monomial $Q_{ab}$ that is fillable by multiplying by $q_{ab}$, with
replicas $a$ and $b$ each appearing only once in $Q_{ab}$. Then
we can write
$$\langle Q_{ab} \rangle_t
= \langle
\sum_{ij}(\sigma_i^a\sigma_j^b/N^2)Q_{ij}(\sigma)\rangle_t$$ where
$Q_{ij}$ contains all factors that do not depend on replicas $a$ or $b$.
Factorizing the state $\Omega_t$ we obtain
\begin{eqnarray*} \langle
Q_{ab} \rangle_t =
\frac{1}{N^2}\mathbb{E}\Big( \sum_{ij}
\omega_t(\sigma_i^a)\omega_t(\sigma_j^b)\Omega_t(Q_{ij}) \Big).
\end{eqnarray*}
Now rewrite the last expression for $t=1$: by
applying the gauge transformation $\sigma_i \rightarrow \sigma_i\sigma_{N+1}$,
the states acting on the replicas $a$ and $b$ are
$\omega_{t=1}(\sigma_i^a)\rightarrow
\omega(\sigma_i^a\sigma_{N+1}^a)+ O(N^{-1})$ while the remaining
product state $\Omega_t$ continues to act on a even number of
occurrences of each replica and is not modified (in a manner directly
analogous to the robustness of averages of filled OCFs).
Putting all the replicas back into a single product
state, we have: \be
\omega(\sigma_i^a\sigma_{N+1}^a)\omega(\sigma_i^b\sigma_{N+1}^b)\Omega(Q_{ij})
= \Omega(\sigma_i^a\sigma_j^b\sigma_{N+1}^a\sigma_{N+1}^bQ_{ij}).
\ee Now the index $N+1$ can be replaced by a dummy index $k$ that is
averaged according to $1=N^{-1}\sum_{k=1}^N$; this does not change the
result except for $O(N^{-1})$ corrections -- from values of $k$ that
coincide with $i$, $j$, or further summation indices in $Q_{ij}$ --
that vanish in the thermodynamic limit. Since $N^{-1}\sum_{k=1}^N
\sigma_{k}^a\sigma_{k}^b = q_{ab}$, this gives the desired result. $\Box$

\subsection{The free energy decomposition}


In this subsection we want to show that the free energy density can be
written in terms of an ``energy-like'' contribution and an
``entropy-like'' one. As a consequence of this decomposition, and
given that we have previously investigated the constraints deriving from the energy-like term we will then restrict our investigation to the entropy-like contribution, which (as we are going to show) is encoded in the cavity function.

It is in fact always possible, via the fundamental theorem of calculus, to relate the free energy to its derivative
with respect to a chosen parameter, here the connectivity $\alpha$. Clearly the result is a relation between the free energy and its $\alpha$-derivative where, interestingly, the missing term is exactly the
cavity function. In the thermodynamic limit this decomposition takes the
form
\begin{equation}\label{sumrule}
P(\alpha,\beta) + \alpha \partial_\alpha P(\alpha,\beta) = \ln
2 + \Psi(\alpha,\beta,t=1).
\end{equation}
We emphasize that the equation above, which we are going to
prove using continuity and the fundamental theorem of calculus,
can be thought of as a generalized thermodynamic definition of the
free energy. In this approach, the cavity function naturally acts
as the thermodynamic entropy, which is why its investigation suggested the title of the paper.

To see briefly how (\ref{sumrule}) arises, let us write down the
partition function of a system of $N+1$
spins at connectivity $\alpha^*=\alpha(N+1)/N$, using the decomposition
(\ref{step}) of the relevant Hamiltonian:
\begin{eqnarray*}
Z_{N+1}(\alpha^*,\beta) &=& e^{\beta \sum_{\nu=1}^{P_{\alpha/N}}
J''_\nu} \sum_{\sigma,\,\sigma_{N+1}= \pm 1} e^{\beta
\sum_{\nu=1}^{P_{\alpha N}}J_{\nu}\sigma_{i_\nu}\sigma_{j_{\nu}}
+\beta\sum_{\nu=1}^{P_{2\alpha}} J'_\nu
\sigma_{i'_\nu}\sigma_{N+1}}
\\
&=& e^{\beta \sum_{\nu=1}^{P_{\alpha/N}} J''_\nu}
\sum_{\sigma,\,\sigma_{N+1}= \pm 1} e^{\beta
\sum_{\nu=1}^{P_{\alpha N}}J_{\nu}\sigma_{i_\nu}\sigma_{j_{\nu}}
+\beta\sum_{\nu=1}^{P_{2\alpha}} J'_\nu \sigma_{i'_\nu}}
\\
&=&2e^{\beta \sum_{\nu=1}^{P_{\alpha/N}} J''_\nu} \sum_{\sigma}
e^{-\beta H_N(\sigma,\alpha;\mathcal{J}) +\beta\sum_{\nu=1}^{P_{2\alpha}}
J'_\nu \sigma_{i'_\nu}},
\end{eqnarray*}
where in going from the first to the second line we have gauge
transformed $\sigma_i \to \sigma_i \sigma_{N+1}$.
%
Multiplying and dividing by
 $Z_N(\alpha,\beta)$ and taking logs we get:
\begin{eqnarray*}
\ln Z_{N+1}(\alpha^*,\beta)= \ln 2 + \beta
\sum_{\nu=1}^{P_{\alpha/N}} J''_\nu +\ln Z_N(\alpha,\beta)
+\ln\omega(e^{\beta\sum_{\nu=1}^{P_{2\alpha}}
J'_\nu\sigma_{i'_{\nu}}}). \nonumber
\end{eqnarray*}
Averaging over the disorder removes the second term and transforms
the last one into the cavity function at $t=1$. Rearranging
slightly, the result reads
\[
[\mathbb{E}\ln Z_{N+1}(\alpha^*,\beta)-\mathbb{E}\ln
Z_{N+1}(\alpha,\beta)]+ [\mathbb{E}\ln Z_{N+1}(\alpha,\beta)
-\mathbb{E}\ln Z_N (\alpha,\beta)] =\ln 2 + \Psi_N(\alpha,\beta,t=1).
\]
Now $\alpha^*-\alpha=\alpha/N$ becomes small as $N$ grows so we can
Taylor expand the first difference on the l.h.s.\ as
$(\alpha/N)\partial_\alpha\mathbb{E} \ln
Z_{N+1}(\alpha,\beta)+O(1/N)$. The second difference on the
l.h.s., on the other hand, gives the pressure as $N\to\infty$, and
so the decomposition (\ref{sumrule}) follows in the limit.
%

\section{Identities from ``direct'' stochastic stability}\label{barra}

Now that the theoretical framework has been outlined, we can turn to
the polynomial identities themselves.
First, in this section, we review the application of a standard approach from fully
connected models to the diluted case \cite{barra1}. To get constraints on averages of overlap polynomials
from this method, one needs to heuristically separate term in a power
series. In the next section, after deriving the general form of the
required cavity streaming equation, we present a modification of
this approach that automatically provides such a separation.
We stress that, in the original SK contest, constraints are usually derived in $\beta$-average \cite{barra1,contucci,gg}, namely one can prove that they are zero in the thermodynamic limit whenever one takes an average over a (however small) $\beta$ interval but not point by point. Here the same results are obtained for the diluted counterpart by tuning $\alpha$ instead of $\beta$, hence constraints are obtained  in  $\alpha$-average.


The standard approach referred to above
studies the family of linear polynomial constraints (identities) on the
distribution of the overlaps which can be obtained by perturbing the
original Gibbs measure defined by the Hamiltonian  (\ref{ham}) with a
random term suggested by the cavity technique, where we use as
a probe for stochastic stability a generalization of the random perturbation given by the connectivity shift from Eq.~(\ref{step}).

Specifically, we consider the quenched expectation of a generic function of
$s$ replicas, with respect to a perturbed measure
defined by the following Boltzmann factor
\begin{equation}\label{pesi}
B(\alpha,\beta,\alpha^{\prime},\beta^{\prime},t) =  \exp\left(-\beta H_{N}(\sigma;\alpha;\mathcal{J})+\beta^{\prime}
    \sum_{\nu=1}^{P_{2\alpha^{\prime} t}}
    \tilde{J}'_{\nu}\sigma_{i'_{\nu}}\right)\ ,
\end{equation}
whose use will be indicated with a subscript $\alpha',\beta',t$ in the
expectations $\Omega_{\alpha',\beta',t}$ and $\langle
. \rangle_{\alpha',\beta',t}$. In this way the
stochastic perturbation coincides with the one from the cavity
technique if we choose $t=1$ and $\alpha'=\alpha$, $\beta'=\beta$,
while for $t=0$ we recover the unperturbed Gibbs measure.

Loosely speaking, after a gauge transformation one can interpret
Eq.~(\ref{pesi}) as the Boltzmann factor of a system of $N+1$ spins,
as in the decomposition (\ref{acca1}). The generalization consists in
letting the additional spin $\sigma_{N+1}$ experience a different
inverse temperature, $\beta'$ rather than $\beta$, and similarly
allow for its connectivity to the other $N$ spins to be set by a
parameter $\alpha'\neq \alpha$. The new variables $\beta^{\prime}$and
$\alpha^{\prime}$ have been introduced for mathematical convenience;
in the end, we will then evaluate everything at $\beta^{\prime}=\beta$
and  $\alpha^{\prime}=\alpha$.

In the following equations we will require powers of
$\tanh(\beta'J)$. Abbreviating $\theta =
\tanh(\beta^{\prime})$ and exploiting that $J=\pm 1$, one has
$\tanh^{2n}(\beta^{\prime} J)=\theta^{2n}$ and
$\tanh^{2n+1}(\beta^{\prime}J)=J\theta^{2n+1}$
$\forall\  n \in \mathbb{N}$.

To see how one can attempt to obtain constraints from the above
generalized stochastic perturbation, consider a generic function $F_s$
of $s$ replicas. Its change with $t$ is given by a ``streaming
equation'' of the following form:
  \begin{multline}\label{stream}
   \partial_t\langle F_s \rangle_{\alpha',\beta',t} =
    -2\alpha^{\prime}\langle F_s
    \rangle_{\alpha',\beta',t} +2\alpha^{\prime} \mathbb{E}\biggl[
\Omega_{\alpha',\beta',t}\biggl(    F_s \{ 1 + J\sum_{a}
    \sigma^{a}_{i_{1}}\theta
    + \sum_{a < b}
    \sigma^{a}_{i_{1}}\sigma^{b}_{i_{1}} \theta^2 \\
    {}+ J\sum_{a < b < c}
    \sigma^{a}_{i_{1}}\sigma^{b}_{i_{1}}\sigma^{c}_{i_{1}}
    \theta^3 + \cdots \} \{ 1 - s
    J\theta \omega_{\alpha',\beta',t}(\sigma) + \frac{s(s+1)}{2!}\theta^2
    \omega_{\alpha',\beta',t}^{2}(\sigma)\\
    {}-\frac{s(s+1)(s+2)}{3!}J\theta^3\omega_{\alpha',\beta',t}^{3}(\sigma)
    + \cdots \}\biggr)\biggr]\ .
  \end{multline}
Here the replica indices $a,b,c$ all run from 1 to $s$. The above
result can be checked by direct calculation and is shown for instance
in \cite{barra4}. We will derive a more general form below.

If one chooses for $F_s$ a function whose average does
not depend on $t$ (for instance a filled OCF, which is robust),
%
the left hand side of (\ref{stream}) is
zero. In other words, one uses as the generator of constraints on the distribution of the overlaps the robustness property
  $$
  \lim_{N\to\infty}
  \partial_t \langle F_s \rangle_{\alpha,\beta,t} = 0
  $$
where $F_s$ is filled and $\alpha^{\prime}=\alpha$, $\beta^{\prime}=\beta$.

For $F_s$ as above, one has on the r.h.s.\ of (\ref{stream})
averages of fillable OCFs. The streaming equation is a power series
in $\theta$ (and hence in $\beta^{\prime}$). Heuristically, driven by the critical behavior of the OCF (as the general multi-overlap $q^2_{2n}$ scales as $(2\alpha\theta-1)^{2n}$ \cite{barra2}),
one can
argue that all coefficients of this power series should vanish if the
l.h.s.\ of (\ref{stream}) vanishes. However, one has to bear in mind
that, before setting $\alpha'=\alpha$ and $\beta'=\beta$, the
averages on the r.h.s.\ will still be dependent on $\beta'$. The r.h.s.\ is not, therefore, a standard power series with constant coefficients.

The simplest example of this reasoning is provided by $F_s=q_{12}^{2}$
with $s=2$. The streaming equation is
\[
  \lim_{N\to\infty}
  \partial_{t}\langle q_{12}^2 \rangle_{\alpha',\beta',t} =
  \lim_{N\to\infty}
  \langle q_{12}^3 - 4 q_{12}^2q_{23} + 3 q^2_{12}q_{34}
  \rangle_{\alpha',\beta',t}\theta^{2} +O(\theta^{4})=
0
\]
If one now assumes that the coefficients of the powers of
$\theta$ are separately zero, then by setting $t=1$, $\alpha'=\alpha$ and
$\beta'=\beta$ one can transform the fillable average into a filled
one to obtain
\[
\lim_{N\to\infty}  \langle q_{12}^4 - 4 q_{12}^2q_{23}^2 + 3 q_{12}^2
  q_{34}^2 \rangle=0
\]
This is the well-known Aizenman-Contucci
relation.


Choosing instead instead $F_s=q^{2}_{1234}$ ($s=4$), and limiting
the expansion to the first two orders $\theta$ of the
streaming equation, one obtains (again for $N\to\infty$)
\begin{multline*}
\partial_{t}\langle q_{1234}^2 \rangle_{\alpha',\beta',t} =
  \theta^2 \langle 3q_{1234}^2q_{12} - 8q_{1234}^2q_{15} + 5
  q_{1234}^2q_{56}\rangle_{\alpha',\beta',t} \\
  {} + \theta^4\langle q_{1234}^3-16q_{1234}^2q_{1235}+60q_{1234}^2q_{1256}
  -80q_{1234}^2q_{1567}+35q^2_{1234}q_{5678}\rangle_{\alpha',\beta',t}
  +O(\theta^{6})=0
\end{multline*}
We re-emphasize that one cannot deduce that each term in this
expansion vanishes separately, as the Boltzmann factor inside the averages is a function of $\beta^{\prime}$ and hence $\theta$, and therefore so are the averages.
If one nevertheless proceeds and sets the coefficient of the
second power of $\theta$ to zero, one obtains at $\alpha'=\alpha$, $\beta'=\beta$ and $t=1$ a relation between filled OCFs:
$$
\langle q_{1234}^2q_{15}^2
\rangle = \frac{3}{8}\langle q_{1234}^2 q_{12}^2\rangle + \frac{5}{8}
\langle q_{1234}^2q_{56}^2 \rangle\ ,
$$
and similarly from the fourth order in $\theta$
$$
\langle q_{1234}^4
\rangle = \langle 16q_{1234}^2q_{1235}^2 -60q_{1234}^2q_{1256}^2
+80q_{1234}^2q_{1567}^2 -35q^2_{1234}q^2_{5678} \rangle\ .
$$
These relations are in perfect agreement with previous investigations \cite{franz2} and generalize the standard identities to diluted systems. Clearly when the connectivity diverges the multi-overlaps go to zero and the equations then no longer provide any non-trivial information.

We note briefly here that a similar heuristic step, where coefficients of a power series are taken to be zero even though they initially depend on the variable that one is expanding in, is used implicitly in the derivation of the AC-like identities in~\cite{franz2}. The same comment applies to the arguments leading to the GG-like relations in~\cite{franz1}.

\section{``Reversing'' stochastic stability}

\subsection{General streaming equation}

To go beyond the heuristic arguments reviewed in the previous section, we need to write down first the general form of the streaming equation (\ref{stream}). The $t$-dependence in the Boltzmann factor (\ref{pesi}) arises only from the Poisson variable $P_{2\alpha't}$. For a generic function of such a variable, leaving off the factor $2\alpha'$ initially, one has
$
\mathbb{E}f(P_t) = e^{-t}\sum_{k=0}^\infty f(k) t^k/k!
$
and so
\[
\partial_t \mathbb{E}f(P_t) = -e^{-t}\sum_{k=0}^\infty f(k) t^k/k! + e^{-t} \sum_{k=1}^\infty f(k) t^{k-1}/(k-1)! = -\mathbb{E}f(P_t) + \mathbb{E}f(1+P_t)
\]
Applying this to $\langle F_s\rangle_{\alpha',\beta',t}$ gives
\[
\frac{\partial_t \langle F_s\rangle_{\alpha',\beta',t}}{2\alpha'} = - \langle F_s\rangle_{\alpha',\beta',t} +
\mathbb{E} \frac{\Omega_{\alpha',\beta',t}(F_s e^{\beta' J \sum_{a=1}^s\sigma_i^a})}{\Omega_{\alpha',\beta',t}(e^{\beta' J \sum_{a=1}^s\sigma_i^a})}
\]
where $J$ is the random interaction strength for the additional coupling and $i$, drawn uniformly from $\{1,\ldots,N\}$, is the associated spin index. The resulting exponential can be written as $\prod_{a=1}^s[\cosh(\beta') +\sinh(\beta')J\sigma_i^a]$ and cancelling the common factor $[\cosh(\beta')]^s$ yields
\[
\frac{\partial_t \langle F_s\rangle_{\alpha',\beta',t}}{2\alpha'} + \langle F_s\rangle_{\alpha',\beta',t} =
\mathbb{E} \frac{\Omega_{\alpha',\beta',t}(F_s \prod_{a=1}^s[1+J\theta\sigma_i^a])}
{\Omega_{\alpha',\beta',t}(\prod_{a=1}^s[1+J\theta\sigma_i^a])}
\]
where $\theta=\tanh(\beta')$ as before. The denominator factorizes across replicas, giving a factor $[1+J\theta\Omega_{\alpha',\beta',t}(\sigma_i)]^{-s}$. One expands this and also the numerator in powers of $\theta$ to get for the r.h.s.\ of the last equation
\[
\mathbb{E} \Omega_{\alpha',\beta',t}\left(F_s
\sum_{k=0}^s(J\theta)^k \sum_{1\leq a_1<\ldots<a_k\leq s}
\sigma_i^{a_1} \cdots \sigma_i^{a_k}\right) \sum_{l=0}^\infty
\frac{(s+l-1)!}{l!(s-1)!}(-1)^l (J\theta)^l [\Omega_{\alpha',\beta',t}(\sigma_i)]^l
\]
To cast each term as an average over a replicated measure again, one can write $[\Omega_{\alpha',\beta',t}(\sigma_i)]^l
 = \Omega_{\alpha',\beta',t}(\sigma_i^{s+1}\cdots\sigma_i^{s+l})$. This gives as the general streaming equation
\be\nonumber
\frac{\partial_t \langle F_s\rangle}{2\alpha'} = \left\langle F_s\left[-1+
\sum_{k=0}^s(J\theta)^k \sum_{1\leq a_1<\ldots<a_k\leq s}
\sigma_i^{a_1} \cdots \sigma_i^{a_k} \sum_{l=0}^\infty
\frac{(s+l-1)!}{l!(s-1)!}(-1)^l (J\theta)^l \sigma_i^{s+1}\cdots
\sigma_i^{s+l}\right]\right\rangle
\ee
where for brevity we have dropped the subscript $\alpha',\beta',t$
on all averages.
Now we gather terms according to equal powers $m=k+l$ of $\theta$, and
note that the $m=0$ term cancels with the $-1$:
\be\nonumber
\frac{\partial_t \langle F_s\rangle}{2\alpha'} = \left\langle
F_s\sum_{m=1}^\infty (J\theta)^{m}
\sum_{l=0}^{m} \sum_{1\leq a_1<\ldots<a_{m-l}\leq s}
\sigma_i^{a_1} \cdots \sigma_i^{a_{m-l}}
\frac{(s+l-1)!}{l!(s-1)!}(-1)^{l} \sigma_i^{s+1}\cdots
\sigma_i^{s+l}\right\rangle.
\ee
Here and below it is understood that the sum over $a_1$, \ldots, $a_{m-l}$
vanishes when $m-l>s$, because it is then not possible to satisfy the
constraint $1\leq a_1<\ldots<a_{m-l}\leq s$.

Performing now the expectation over $J$ cancels all odd orders $m$, and
the expectation over $i$ produces a multi-overlap, giving
\be\nonumber
\frac{\partial_t \langle F_s\rangle}{2\alpha'} = \left\langle
F_s\sum_{m=2,4,\ldots} \theta^{m}
\sum_{l=0}^{m} \frac{(s+l-1)!}{l!(s-1)!}(-1)^{l}
\sum_{1\leq a_1<\ldots<a_{m-l}\leq s}
q_{a_1\ldots a_{m-l},s+1\ldots s+l}\right\rangle.
\ee
To state this result in a compact form, we define
%
\be\nonumber
C^{(m,n)}_s =
\sum_{l=0}^{m} \frac{(s+l-1)!}{l!(s-1)!}(-1)^{l}
\sum_{1\leq a_1<\ldots<a_{m-l}\leq s}
q^n_{a_1\ldots a_{m-l},s+1\ldots s+l}.
%
\label{C_def}
\ee
The superscript $m$ indicates how many replicas are involved in each of the overlaps in this expression. Each overlap is taken to the power $n$, a generalization which will be useful shortly. Then after re-instating the subscripts on the averages, the streaming equation can be written simply as
\be
\frac{\partial_t \langle F_s\rangle_{\alpha',\beta',t}}{2\alpha'} = \left\langle
F_s\sum_{m=2,4,\ldots} \theta^{m} C_s^{(m,1)}\right\rangle_{\alpha',\beta',t}.
\label{stream_general}
\ee
%

We can now state the arguments of the previous sections in more general form. If $F_s$ is filled then the derivative on the l.h.s.\ of (\ref{stream_general}) must vanish. If one evaluates at
$\alpha'=\alpha$, $\beta'=\beta$, $t=1$ and for $N\to\infty$, all the overlaps in $C_s^{(m)}$
become squared -- the fillable averages become filled -- so each factor $C^{(m,1)}_s$ turns into a ``higher order AC factor'' $C_s^{(m,2)}$.
The identities from directed stochastic stability are therefore (for filled $F_s$ and in the limit $N\to\infty$)
\be\nonumber
\left\langle
F_s\sum_{\gamma\o=2,4,\ldots} \theta^{m}
C_s^{(m,2)}\right\rangle = 0.
\ee
As explained above, one cannot necessarily separate the different orders in $\theta$ in this result. This is possible by ``reversing'' the approach, as we will now see.

\subsection{Identities from reversed stochastic stability}

Progress in demonstrating that each term of the above expression must vanish separately, i.e.\
$\left\langle F_s C_s^{(m,2)}\right\rangle = 0$ for each $s$ and even $m$, can be made by considering not $t=1$ but derivatives at $t=0$ and assuming that, in  general, stochastic stability holds \cite{pierluz,contucci}. Let us take a generic overlap polynomial $F_s$, and consider as above the ``smooth cavity'' perturbation with a modified temperature $\beta'$ and corresponding $\theta=\tanh(\beta')$. Then the first derivative w.r.t.\ $t$ is (writing
$m=2n$ now)
\be\nonumber
\frac{\partial_t \langle F_s\rangle_{\alpha',\beta',t}}{2\alpha'} =
\sum_{n=1}^\infty
\theta{}^{2n} \left\langle F_s  C_s^{(2n,1)}\right\rangle_{\alpha',\beta',t}.
\ee
Pulling the infinite sum out of the expectation and differentiating again (assuming that we can interchange
differentiation with the infinite sum over $n$) we get
\be
\frac{\partial^2_t \langle F_s\rangle_{\alpha',\beta',t}}{(2\alpha')^2} =
\sum_{n=1}^\infty
\sum_{m=1}^\infty \theta^{2(m+n)}
\left\langle F_s  C_s^{(2n,1)}C_{s+2n}^{(2m,1)}\right\rangle_{\alpha',\beta',t}.
\label{t0_aux}
\ee
From now on, take $F_s$ to be filled and set $t=0$ instead of $t=1$: we can then drop
the subscripts on the average on the r.h.s.\ because at $t=0$ we have
an unperturbed Boltzmann average. Then because of gauge invariance of
this unperturbed state, all terms with $m\neq n$ give vanishing
averages: $F_s$ is filled, $C_s^{(2n,1)}$ consists of a sum of $n$-th
order overlaps of $n$ distinct replicas, and $C_{s+2n}^{(2m,1)}$ of a
sum of $m$-th order overlaps of $m$ distinct replicas. At least
$|m-n|\geq 1$ replicas therefore occur an odd number of times in all
possible combinations of terms. Hence we can collapse the sum to
\be
\left.\frac{\partial_t^2 \langle
    F_s\rangle_{\alpha',\beta',t}}{(2\alpha')^2}\right|_{t=0} =
\sum_{n=1}^\infty \theta^{2n}
\left\langle F_s  C_s^{(2n,1)}C_{s+2n}^{(2n,1)}\right\rangle.
\label{t0_aux2}
\ee
One can simplify further: $C_s^{(2n,1)}$ is as function only of replicas
$1,\ldots, s+2n$. This means that in the sum (\ref{C_def}) defining
$C_{s+2n}^{(2n,1)}$, only the term with $l=0$ can give non-vanishing
averages, because all other terms depend on some of the replicas
$s+2n+1,\ldots, s+4n$ and these would remain unpaired. In other words,
\be\nonumber
\left\langle F_s  C_s^{(2n,1)}C_{s+2n}^{(2n,1)}\right\rangle =
\left\langle F_s  C_s^{(2n,1)}
\sum_{1\leq b_1<\ldots<b_{2n} \leq s+2n}
q_{b_1\ldots b_{2n}} \right\rangle.
\ee
But now for each term $q_{a_1,\ldots,a_{2n}}$ in $C_s^{(2n,1)}$ (with
$1\leq a_1<\ldots<a_{2n}\leq s+2n$) there is exactly one entry in the
sum over $1\leq b_1<\ldots<b_{2n}\leq s+2n$ which fills this term and
gives a nonzero average, namely $b_1=a_1$, \ldots, $b_{2n}=a_{2n}$.
The multiplication by the sum over $b_1,\ldots,b_{2n}$ therefore just
has the effect of squaring all the overlaps in $C_s^{(2n,1)}$ and we get
\be\nonumber
\left\langle F_s  C_s^{(2n,1)}C_{s+2n}^{(2n,1)}\right\rangle =
\left\langle F_s  C_s^{(2n,2)}\right\rangle.
\ee
Inserting into (\ref{t0_aux2}) gives then
\be\nonumber
\left.\frac{\partial_t^2 \langle
    F_s\rangle_{\alpha',\beta',t}}{(2\alpha')^2}\right|_{t=0} =
\sum_{n=1}^\infty \theta{}^{2n}
\left\langle F_s  C_s^{(2n,2)}\right\rangle.
\ee
%

Now we assume that we have stochastic stability for generic
$\theta$~\cite{pierluz,contucci}
and $t$, i.e.\ that $\langle F_s\rangle_{\alpha',\beta',t}$ is independent of
$\beta'$ and $t$ in the limit $N\to\infty$. Then the last expression vanishes and hence so must all coefficients of different powers of $\theta$. We thus
obtain the relations (for $s\geq 2$ and $n\geq 1$)
\be
\lim_{N\to\infty} \langle F_s C_s^{(2n,2)}\rangle = 0.
\label{general_identity}
\ee
This includes for $n=1$ all the standard SK AC-identities, but also
all their higher-order generalizations.

We now compare with closely related identities obtained by de Sanctis and Franz~\cite{franz1}, and Franz, Leone and Toninelli \cite{franz2}. It is easy to check that the identities obtained in these papers can be written in our notation as
\[
\lim_{N\to\infty} \langle q_{1\ldots s}^{2r} C_s^{(2n,2p)}\rangle = 0.
\]
for arbitrary positive integers $r$ and $p$. The identities (\ref{general_identity}) relate to $p=1$ but are then rather more general because they allow arbitrary filled overlap polynomials for $F_s$. For example our identities for $F_3=q_{12} q_{23} q_{13}$ are not contained in the set of identities from~\cite{franz1,franz2}.

\subsection{Generalization to fields with multiple spins}

The generalization of the identities (\ref{general_identity}) to exponents greater than two can be achieved relatively simply by a standard approach, allowing not just fields but $p$-spin interactions in the perturbing term.
%
%
The perturbation term from (\ref{pesi}) would then be generalized to
\[
\beta^{\prime} \sum_{\nu=1}^{P_{2\alpha^{\prime} t}} \tilde{J'}_{\nu}\sigma_{i^1_{\nu}}
\cdots \sigma_{i^{p-1}_{\nu}},
\]
and reduces to the latter for $p=2$ as it should.

In the streaming equation w.r.t.\ $t$, factors like $\sigma_i^a$ are then consistently replaced by $\sigma_{i^1}^a \cdots \sigma_{i^{p-1}}^a$, and accordingly one obtains in the end
\be
\frac{\partial_t \langle F_s\rangle_{\alpha',\beta',t}}{2\alpha'} = \left\langle
F_s\sum_{m=2,4,\ldots} \theta^{m} C_s^{(m,p-1)}\right\rangle_{\alpha',\beta',t}.
\label{general_1st}
\ee
The calculation of the second derivative at $t=0$ generalizes in the same way, giving
\begin{equation}
\left.\frac{\partial_t^2 \langle
    F_s\rangle_{\alpha',\beta',t}}{(2\alpha')^2}\right|_{t=0} =
\sum_{n=1}^\infty \theta{}^{2n}
\left\langle F_s  C_s^{(2n,2(p-1))}\right\rangle.
\label{general_2nd}
\end{equation}
Based again on the assumption of stochastic stability under a small perturbation of the Boltzmann measure caused by the introduction of $O(N^0)$ random $(p-1)$-spin interaction terms into a system of $N$ spins, one then deduces the identities
\be
\lim_{N\to\infty} \langle F_s C_s^{(2n,2(p-1))}\rangle = 0.
\label{most_general_identity}
\ee
With the overlap exponent now generalized from $2$ to $2(p-1)$, these form a strict superset of the identities from~\cite{franz2,franz1}.

We have been somewhat casual above in not distinguishing between odd and even order $p-1$ of the perturbing Hamiltonian. Specifically, the reasoning that leads to the simplified form of the second derivative~(\ref{general_2nd}) works, by analogy with the case $p=2$, only when $p-1$ is odd. For even $p-1$ a more complicated expression analogous to (\ref{t0_aux}) would result. However, in this case one can exploit the vanishing of the first derivative (\ref{general_1st}). Writing $p-1=2(p'-1)$ because $p-1$ is even, one then obtains again the identities (\ref{most_general_identity}), with $p$ replaced by $p'$ which is now an arbitary integer.

While we have assumed throughout Poissonian graphs, where each pair interaction is present independently of the others, one would expect that the identities (\ref{most_general_identity}) hold also for spin glass model on other graphs with finite connectivity. The treatment in~\cite{franz2} is more general in this regard, and it may be possible to adapt the methods used there to generalize~(\ref{most_general_identity}) to this broader range of settings.

Our reasoning leading to the general higher-order AC-identitites (\ref{most_general_identity}) is, like the one in Ref.~\cite{franz2,franz1}, not rigorous. The main assumption here, as in the other papers, is stochastic stability for general values of $\beta'\neq \beta$. More technically, we have also been somewhat cavalier in our treatment of infinite sums, interchanging e.g.\ differentation w.r.t.\ $t$ with summation.

\section{Conclusion}

The phenomenon of full replica symmetry breaking, and its probably
best known consequence, namely ultrametricity, have deep implications
in physics. It is for this reason that the Sherrington-Kirkpatrick
model, which is the fully connected limit of the Viana-Bray diluted
spin glass discussed here, is sometimes described as the harmonic oscillator of complex systems.

Parisi ultrametricity is a strong constraint on overlap probability
distributions, and entails peculiar constraints for averages of
polynomials of these. These linear (in the averages) polynomial
identities were the subject of this paper. They are of interest in
themselves, but also with regards to the question of how far results
from the fully connected mean field framework can be extended to other
scenarios.

Linear identities in diluted spin glasses have already been obtained
with standard techniques, mainly intensivity of the internal energy \cite{franz2}.
In our own work we have shown how they can be embedded
within the framework of random overlap structures, and have analyzed
in particular what identities can be obtained from
the ``energy'' contribution to the free energy in this context \cite{peter}.
In this paper we showed
how to recover all known linear identities by focussing attention on
the entropy instead of the internal energy and then we highlighted how our method allowed us to further enlarge the set of identities.

We reviewed first the results of a classical
stochastic stability analysis, which contains within it, via a gauge
transformation, the physics of the cavity approach. Linear
identities for multi-overlaps can be argued for within this method,
but require a heuristic separation into the different orders in
$\theta=\tanh(\beta')$, where $\beta'$ is the inverse temperature
associated with the stochastic perturbation.

The main contribution of this paper was then to go beyond
this. We started from a general streaming equation describing the effect
of a stochastic perturbation on the Gibbs measure.
Instead of perturbing the Gibbs
measure and then using the thermodynamic limit to make the effect of the
perturbation vanish, we evaluated averages directly in the
unperturbed Gibbs state, using second derivatives with respect to the
perturbation parameter $t$. This gives stronger results, in that
identities from different orders in $\theta$ can be cleanly separated; it also yields a larger number of identities compared to those obtained previously using other techniques. 

Conceptually, it is interesting that this new approach does not
directly exploit the cavity nature of the perturbation, i.e.\ the
mapping to a system of $N+1$ spins when $t=1$ and $\beta'=\beta$. But it
certainly does use the gauge invariance of the unperturbed Boltzmann state.

As we have emphasized, our arguments are not rigorous, requiring as
the key assumption stochastic stability under general perturbations as
well as some more technical conditions. We would hope, however, that
our reasoning might in the future form the basis for a rigorous proof
of all linear polynomial identities in diluted spin glasses.

\section*{Acknowledgements}

AB acknowledges the FIRB grant RBFR08EKEV and Sapienza Universit\`a di Roma for partial financial support.
The authors are grateful to Francesco Guerra, Pierluigi Contucci and Cristian Giardin\`a
for enlightening conversations.

\end{document}